\documentclass[reprint,groupedaddress]{revtex4-1}
\usepackage{graphicx}
\usepackage{epstopdf}
\usepackage{bm, amsmath, amssymb}
\usepackage{float}
\usepackage{placeins}
\usepackage{color}
\usepackage{siunitx}
\usepackage[T1]{fontenc}

\newcommand{\bfsc}[1]{\textbf{\textsc{#1}}}

\begin{document}

\title{3-Wave Mixing Josephson Dipole Element}

\author{N. E. Frattini}
\email{nicholas.frattini@yale.edu}
\affiliation{Department of Applied Physics, Yale University, New Haven, CT 06520, USA}
\author{U. Vool}
\affiliation{Department of Applied Physics, Yale University, New Haven, CT 06520, USA}
\author{S. Shankar}
\affiliation{Department of Applied Physics, Yale University, New Haven, CT 06520, USA}
\author{A. Narla}
\affiliation{Department of Applied Physics, Yale University, New Haven, CT 06520, USA}
\author{K. M. Sliwa}
\affiliation{Department of Applied Physics, Yale University, New Haven, CT 06520, USA}
\author{M. H. Devoret}
\email{michel.devoret@yale.edu}
\affiliation{Department of Applied Physics, Yale University, New Haven, CT 06520, USA}

\date{\today}

\begin{abstract}
Parametric conversion and amplification based on three-wave mixing are powerful primitives for efficient quantum operations. For superconducting qubits, such operations can be realized with a quadrupole Josephson junction element, the Josephson Ring Modulator (JRM), which behaves as a loss-less three-wave mixer. However, combining multiple quadrupole elements is a difficult task so it would be advantageous to have a three-wave dipole element that could be tessellated for increased power handling and/or information throughput. Here, we present a dipole circuit element with third-order nonlinearity, which implements three-wave mixing. Experimental results for a non-degenerate amplifier based on the proposed third-order nonlinearity are reported.
\end{abstract}

\maketitle

In quantum devices based on superconductors, Josephson junctions provide a nonlinear interaction between electromagnetic modes which is purely dispersive. However, because the Josephson potential is an even function of the superconducting phase difference $\bm{\varphi}$, this nonlinearity is, to lowest order, of the form $\bm{\varphi}^4$. This Kerr nonlinearity is useful for engineering interactions between modes, but it imparts undesired frequency shifts. This problem is analogous to that generated by $\chi^{(3)}$ media in nonlinear optics~\cite{boyd_nonlinear_2008}. Such frequency shifts become especially problematic as the number of interacting modes, and therefore frequency crowding, increases. An alternative strategy is to use a minimal $\bm{\varphi}^3$ nonlinearity for engineering the same useful interactions between modes while minimizing these unwanted Kerr frequency shifts.

A form of this $\bm{\varphi}^3$ nonlinearity has been realized with four Josephson junctions arranged in a ring threaded by DC magnetic flux. This device, called the Josephson ring modulator (JRM)~\cite{bergeal_phase-preserving_2010, bergeal_analog_2010}, specifically provides a trilinear Hamiltonian term of the form $\bm{\varphi_{\bfsc{x}} \varphi_{\bfsc{y}} \varphi_{\bfsc{z}}}$ between three modes named $X$, $Y$, and $Z$. However, the JRM is a quadrupole element, i.e.~it imposes a current and phase relation between four nodes of a circuit. Thus, a question naturally arises, would it be possible to engineer a a third-order $\bm{\varphi}^3$ nonlinearity  in a dipole device similar to the Josephson junction?

An inherent advantage of such a dipole element is that it is more modular and easier to integrate into complex circuits. In particular, such an element can be tessellated, as with a Josephson junction~\cite{castellanos-beltran_amplification_2008}, for improved power handling capabilities and information throughput. Furthermore, it can provide a three-wave coupling between a superconducting qubit and an oscillator.

In this Letter, we report an experimental realization of a pure $\bm{\varphi}^3$ nonlinear dipole element, demonstrating its performance in a non-degenerate parametric amplifier. Because this dispersive element is asymmetric in the transformation of $\bm{\varphi}\to -\bm{\varphi}$, in contrast with the SQUID~\cite{zimmerman_macroscopic_1966,clarke_squid_2004} and the SLUG~\cite{clarke_superconducting_1966, hover_superconducting_2012}, we have named it the Superconducting Nonlinear Asymmetric Inductive eLement (SNAIL). After the completion of our work, we became aware of a theoretical proposal for a similar element, given in Ref.~\onlinecite{zorin_josephson_2016}.
\color{black}

\begin{figure*}
 \includegraphics[angle = 0, width = \textwidth]{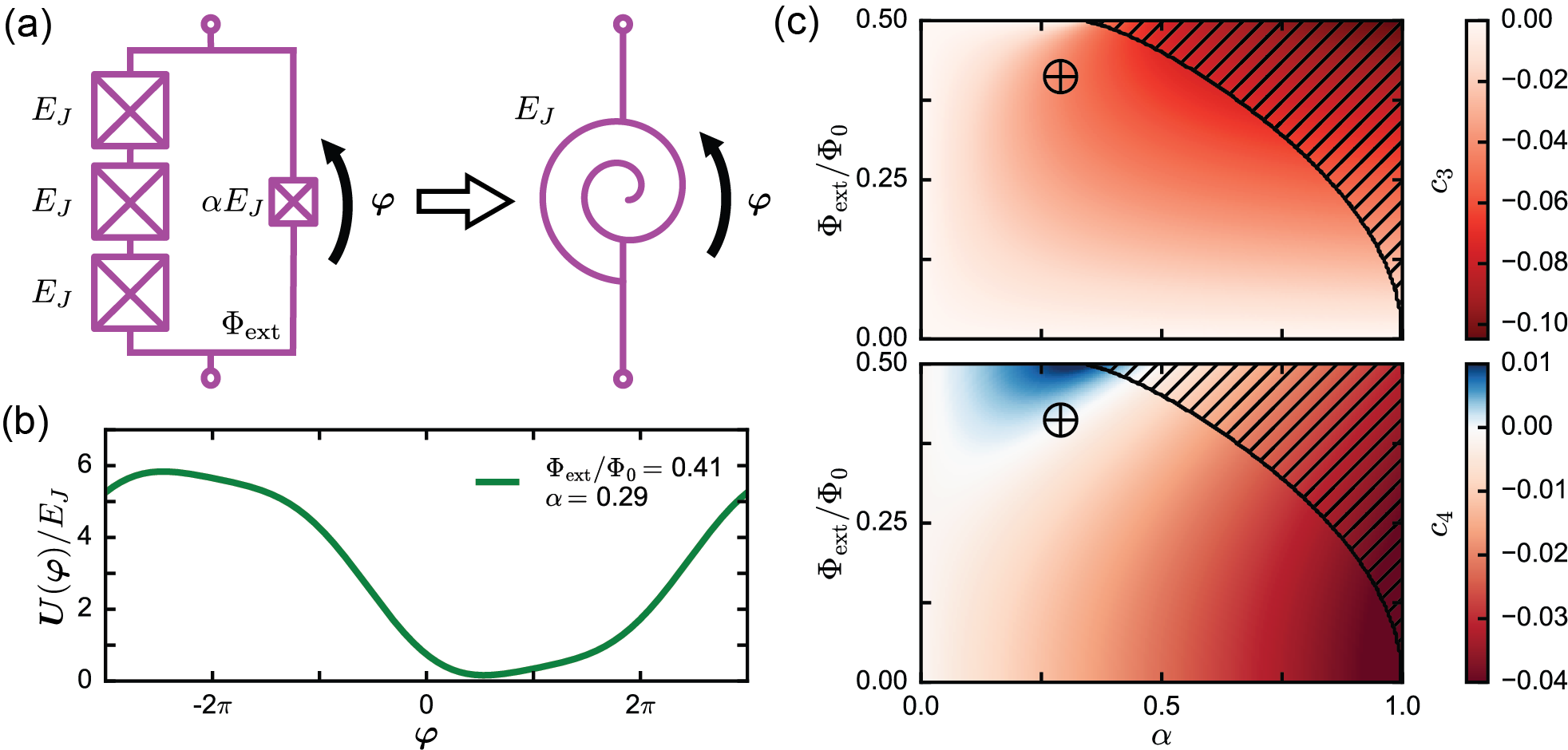}
 \caption{\label{fig1}  (a) Circuit for the Superconducting Nonlinear Asymmetric Inductive eLement (SNAIL) reduced to one degree of freedom $\bm{\varphi}$. The loop of $n=3$ large junctions and one smaller junction (tunneling energies $E_J$ and $\alpha E_J$ respectively) is threaded with an external DC flux $\Phi_{\mathrm{ext}}$. (b) An example SNAIL potential for $\alpha = 0.29$ and $\Phi_{\mathrm{ext}} = \SI{0.41}{\Phi_0}$ that includes third-order nonlinearity without fourth-order nonlinearity. (c) Colormaps of the $(\alpha, \Phi_{\mathrm{ext}})$ parameter space for the third-order (top) and fourth-order (bottom) nonlinear terms show regions where $c_3 \neq 0$ while $c_4 = 0$ (white region in bottom panel (c)). The black-hatched regions in (c) correspond to unwanted hysteretic double well behavior similar to the flux qubit regime. The crosses in (c) mark the set of parameters chosen for (b). This working point is optimal (see text).}
\end{figure*}

\begin{figure}
\includegraphics[angle = 0, width = \columnwidth]{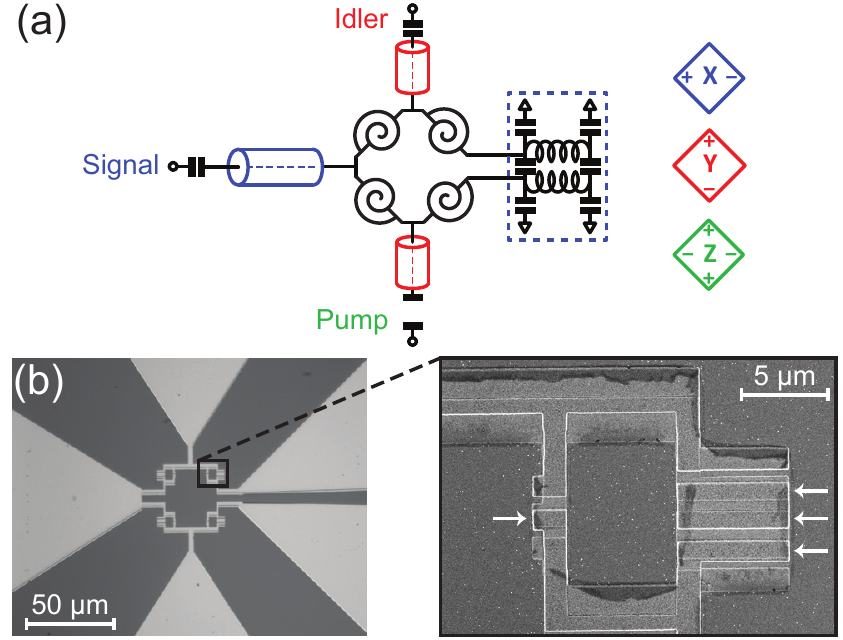}
\caption{\label{fig2}   (a) A circuit design for a non-degenerate three-wave mixing amplifier which uses the SNAIL and is based on the design of the JRM. The elements in the dashed blue box represent a coupled microstrip line whose even mode electrical length is $\lambda_{\rm{S}}/4$. The coupled microstrip is open at DC, and so, unlike the JRM, there is no exterior loop in this design. Depictions of the $X$, $Y$, and $Z$ eigenmodes of the device are shown by the diamonds on the right. The $+$ and $-$ signs refer to the relative phase of the voltage across the four SNAILs. In (b), we show an optical image of the four SNAIL mixing structure, where the inset shows an individual SNAIL with $n=3$ large junctions. The arrows highlight the four junctions. For discussion of the black veil around the junctions, see text.}
\end{figure}

\begin{figure*}
\includegraphics[angle=0, width = \textwidth]{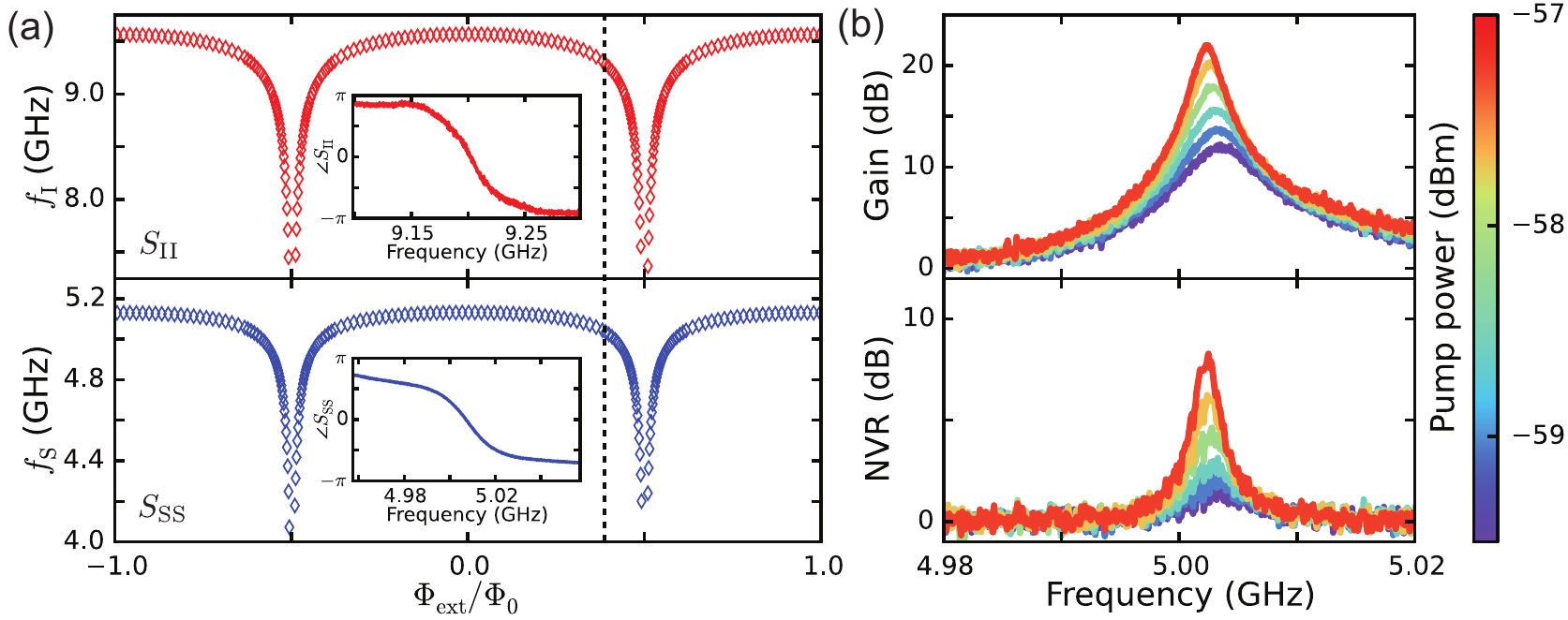}
\caption{\label{fig3}   (a) Resonant frequencies of the Signal (red) and Idler (blue) modes as a function of external applied flux. The inset shows the phase of the reflection coefficient at $\Phi_{\rm{ext}}=0.41 \Phi_0$ (black dashed line), used to fit resonant frequency at each flux. Biasing at the dashed line in (a) and applying a strong microwave pump tone at $f_{\rm{P}}=f_{\rm{S}}+f_{\rm{I}}$, the top panel of (b) shows the reflection gain off the Idler port. Gain increased and bandwidth decreased as a function of pump power (indicated by color). As a proxy for noise temperature, the bottom panel of (b) shows the Noise Visibility Ratio ($\rm{NVR} = P_{\rm{on}}/P_{\rm{off}}$). The $\rm{NVR}$ of our device was comparable to that of other quantum-limited amplifiers measured in the same system, suggesting the noise performance of the SNAIL is similar to that of the nearly quantum-limited JRM.}
\end{figure*}

\begin{figure}
\includegraphics[angle = 0, width = \columnwidth]{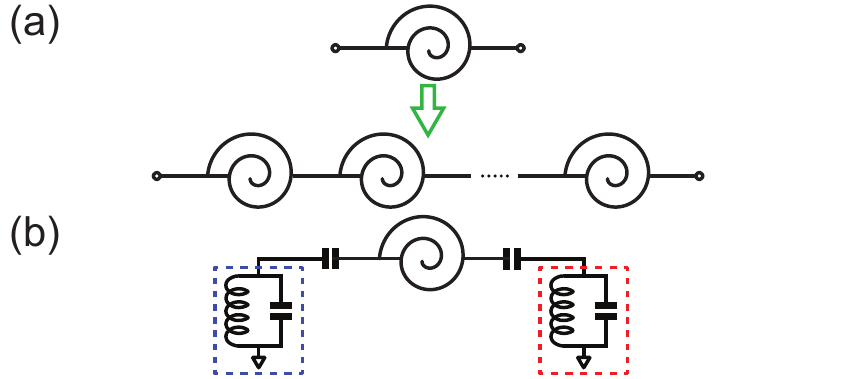}
\caption{\label{fig4}   (a) Replacing a SNAIL with an array of SNAILs will allow for increased dynamic range or bandwidth, similar to the tessellation of DC-SQUIDs in the JPA. The dipolar nature of the SNAIL allows this tesselation for three-wave mixing amplifiers. (b) An example of a circuit in which the SNAIL is used as a tunable coupling device between two modes (red and blue boxes).}
\end{figure}

Fig.~1a shows the circuit we propose to achieve cubic nonlinearity in a dipole circuit element. The SNAIL consists of a superconducting loop of $n$ large Josephson junctions and a single smaller junction (tunneling energies $E_J$ and $\alpha E_J$ respectively), which we thread with DC magnetic flux $\Phi_{\rm{ext}}$. The proposed circuit has the inductive energy:
\begin{equation}
\label{AMJcos}
\bm{U_{\mathrm{SNAIL}}}(\bm{\varphi})=-\alpha E_J \cos(\bm{\varphi}) - n E_J \cos\left(\frac{\varphi_{\mathrm{ext}}-\bm{\varphi}}{n}\right)
\end{equation}
where $\bm{\varphi}$ is the superconducting phase across the small junction, $\varphi_{\mathrm{ext}} = 2\pi\Phi_{\mathrm{ext}}/\Phi_0$ is the reduced applied magnetic flux, and $\Phi_0 = h/2e$ is the magnetic flux quantum. Note that Eq.~\ref{AMJcos} is only a function of a single degree of freedom $\bm{\varphi}$ as we eliminated the dynamics due to any intra-array modes and considered only common excitations across the array of $n$ junctions. This reduction is valid when $E_J \gg E_C$ for each junction where $E_C = e^2/2C_J$ is the Coulomb charging energy of the junction with capacitance $C_J$, and when $C_0 \ll C_J/n^2$ where $C_0$ is the capacitance to ground of each island between junctions~\cite{matveev_persistent_2002}. We will assume in the following that these conditions are realized.

With $\alpha \sim 0.8$ and $\Phi_{\rm{ext}} \sim \SI{0.5}{\Phi_0}$, this circuit is well known as the flux qubit~\cite{mooij_josephson_1999,wal_quantum_2000}, which has a double-well potential. However, here we propose using a different parameter set to create a potential with a single minimum. Moreover, we can adjust the potential to cancel the fourth-order (Kerr) term while keeping a substantial cubic term for a particular choice of $\alpha$ and $\Phi_{\rm{ext}}$.

Through numerical minimization of Eq.~\ref{AMJcos}, we analyze the SNAIL's mixing capabilities by Taylor expanding about the minimum $\varphi_{\rm{min}}$ to obtain the effective potential for $\bm{\tilde{\varphi}} = \bm{\varphi} - \varphi_{\rm{min}}$:
\begin{equation}
\label{AMJexpan}
\bm{U_{\rm{eff}}}(\bm{\tilde{\varphi}})/E_J=c_2 \bm{\tilde{\varphi}}^2 + c_3 \bm{\tilde{\varphi}}^3 + c_4 \bm{\tilde{\varphi}}^4 + \cdots
\end{equation}
where ($c_2$, $c_3$, $c_4$, $\cdots$) are numerically determinable coefficients whose specific values depend on $n$, $\alpha$ and $\Phi_{\mathrm{ext}}$. Our desire for a pure cubic nonlinearity without any Kerr translates to a $\bm{U_{\rm{eff}}}$ with nonzero $c_3$, but $c_4=0$. Note that the case of $n=1$, corresponding to the SQUID, always gives $c_3 = 0$ since the potential about $\varphi_{\rm{min}}$ is always a pure cosine irrespective of the values of $\alpha$ and $\Phi_{\rm{ext}}$. Interestingly, $\alpha$ and $\Phi_{\rm{ext}}$ are not enough to break the $\bm{\tilde{\varphi}} \to - \bm{\tilde{\varphi}}$ symmetry in this case. Additionally, in the $n\gg 1$ limit, the array behaves as a linear inductance and the potential approaches the fluxonium qubit/RF-SQUID regime~\cite{manucharyan_fluxonium:_2009,masluk_reducing_2012}. Replacing the array with a geometric inductance in an RF-SQUID configuration also achieves the desired nonlinearity~\cite{zorin_josephson_2016}. Experimentally, we choose the smallest $n \geq 2$ that is easy to fabricate, which is $n=3$ for our Dolan bridge process. We specialize all further analysis to the case $n=3$ as depicted in Fig.~1a, but the results are easily extendable to different values of $n$.

For the SNAIL to achieve Kerr-free three-wave mixing, we must optimize on the parameter space of ($\alpha$, $\Phi_{\rm{ext}}$). The top (bottom) panel of Fig.~1c shows $c_3$ ($c_4$) as a function of the ($\alpha$, $\Phi_{\rm{ext}}$) parameter space. Focusing on the top of Fig.~1c, we want to maximize $c_3$ while avoiding any hysteretic double-well behavior, marked by the black-hatched region. This is done by restricting ourselves to $\alpha \lesssim 0.5$. The Kerr-free region of parameter space corresponds to $c_4=0$ (the white region in the bottom panel of Fig.~1c). As such, we choose an optimal set of parameters $\alpha = 0.29$ and $\Phi_{\rm{ext}} = \SI{0.41}{\Phi_0}$, marked by the crosses in Fig.~1c, which maximizes $c_3$ while $c_4 =0$ and the potential has a single minimum. We plot one period of the $\bm{U_{\rm{SNAIL}}}$ potential (Eq.~\ref{AMJcos}) for this optimal set of parameters in Fig.~1b.

In order to experimentally verify our model for the SNAIL as a cubic nonlinearity, we integrate it into a design for a non-degenerate three-wave mixing amplifier. Such an amplifier consists of two spatially and spectrally separate modes that are coupled by a pump applied at the sum of their respective frequencies. Achieving amplification with this pump condition is an explicit signature of a cubic nonlinearity~\cite{boyd_nonlinear_2008}. The design (Fig.~2a) is based on that of the Josephson Ring Modulator (JRM)~\cite{bergeal_analog_2010, bergeal_phase-preserving_2010}, a quadrupole element with four identical Josephson junctions in a superconducting ring. By threading the ring with an external magnetic flux, the JRM achieves the three-wave coupling $\bm{\varphi_{\bfsc{x}} \varphi_{\bfsc{y}} \varphi_{\bfsc{z}}}$ between modes $X$, $Y$, and $Z$. This coupling allows the device to act as a parametric amplifier and frequency converter.

To achieve the same coupling term, we arrange four identical SNAILs into a ring. Like the JRM, this Wheatstone bridge of SNAILs cancels all cubic terms in the Hamiltonian of the device except the desired $\bm{\varphi_{\bfsc{x}} \varphi_{\bfsc{y}} \varphi_{\bfsc{z}}}$. This cancellation assumes that all four SNAILs are identical. Note that inverting the polarity of one SNAIL would destroy this cancellation, since $\bm{U_{\rm{eff}}}$ (Eq.~\ref{AMJexpan}) depends on the sign of $\bm{\tilde{\varphi}}$. While the JRM relies on the flux through the exterior ring to achieve a cubic potential, in our design each SNAIL individually achieves $\bm{\tilde{\varphi}}^3$, and so we can eliminate the exterior DC loop in the ring. This is advantageous as it prevents an extra DC persistent current from biasing the SNAILs, and allows for future tessellation without introducing a prohibitively large DC loop. Moreover, the SNAIL allows flexibility for other applications currently unfeasible with the JRM, such as endowing a single mode with a cubic nonlinearity. Further applications may also prefer the expression of terms such as a $\bm{\varphi_{\bfsc{x}}}^2 \bm{\varphi_{\bfsc{z}}}$ for degenerate three-wave mixing~\cite{yamamoto_flux-driven_2008}, achievable by simply inverting the polarity of two opposite SNAILs in the ring. In this way, the SNAIL allows the circuit designer to separate the size of the loop for flux biasing from the design choice of which third-order terms are expressed and which are canceled.

For enhancement of the nonlinearity and maximal isolation, we designed a resonant embedding structure using microstrip resonators~\cite{abdo_nondegenerate_2013, schackert_practical_2013}, which couple to the $X$, $Y$, and $Z$ modes of the SNAIL ring (Fig.~2a). The highest frequency mode, called the idler, couples to the $Y$ mode of the ring via two $\lambda_{\rm{I}}/4$ sections of standard microstrip transmission line (red cylinders), where $\lambda_{\rm{I}}$ is the wavelength of the idler mode. For the $X$ mode of the ring, called the signal, on one side we use similar coupling through a $\lambda_{\rm{S}}/4$ section of standard microstrip transmission line (blue cylinder). On the other side, however, we use an impedance-matched coupled microstrip transmission line (blue-boxed elements), which breaks the exterior loop at DC but allows RF coupling to $\bm{\varphi_{\bfsc{x}}}$. Gap capacitors couple the signal and idler modes to $\SI{50}{\ohm}$ transmission lines to set the quality factor $Q \sim 150$ for this device. A small coupling capacitor weakly couples our pump port to the $Z$ mode of the ring to allow driving of parametric processes for three-wave mixing.

Using standard Dolan bridge techniques for aluminum Josephson junctions on silicon, we fabricated a ring of four SNAILs according to the above design (Fig.~2b). The inset of Fig.~2b shows an electron micrograph of a single SNAIL element with $n=3$ large junctions (critical currents $I_0 = \SI{7.1}{\micro \ampere}$) and a single smaller junction ($I_0 = \SI{2.0}{\micro \ampere}$). The ratio of junction sizes gave $\alpha = 0.29$ as desired. For the array of large junctions, we utilized a technique developed for the original fluxonium arrays~\cite{manucharyan_fluxonium:_2009, masluk_reducing_2012} that allows for creation of three equivalent junctions using only two Dolan bridges. Characteristics of the standard microstrip embedding structure were similar to past work~\cite{schackert_practical_2013}. The black veil surrounding the junctions in the micrograph results from undesired leftover resist in those locations during deposition.~\footnote{In principle, this undesired material places an upper bound on the internal quality factor of our device, but this bound is significantly higher than the external quality factor. This black veil is likely to be the result of the resist dose associated with our large junctions.}

For experimental characterization, we mounted the device at the base stage of a helium dilution refrigerator with access to both the signal and idler ports. A uniform external magnetic flux $\Phi_{\rm{ext}}$ was applied to each SNAIL by a magnet coil mounted under the sample. Fig.~3a shows the resonant frequency $f_{\rm{S}}$ ($f_{\rm{I}}$) of the signal (idler) mode as a function of $\Phi_{\rm{ext}}$, determined through reflection measurements off of the signal (idler) port (example shown in the inset). Sweeping $\Phi_{\rm{ext}}$ in the opposite direction resulted in the same mode frequencies, implying our device was not hysteretic. The dashed black line corresponds to the calculated optimal flux $\Phi_{\rm{ext}} =\SI{0.41}{\Phi_0}$ for this sample with $\alpha = 0.29$.

Biasing at this optimal flux point, we applied a strong microwave pump tone on the pump port at frequency $f_{\rm{P}}=f_{\rm{S}}+f_{\rm{I}}$ to drive the parametric amplification process. Achieving parametric amplification with this frequency condition provides the explicit signature of three-wave mixing, and hence the presence of a $\varphi^3$ nonlinearity. The top panel of Fig.~3b shows the reflection gain off of the signal port. Increasing the pump power (color) increased the gain and decreased the bandwidth in accordance with the gain-bandwidth product, standard for cavity-based parametric amplifiers~\cite{clerk_introduction_2010}. We also observed a slight shift in frequency for the maximum gain as we increased the pump power. This shift is similar in magnitude to that observed with a JRM. Further study of this shift and its relation to the Kerr nonlinearity~\cite{liu_josephson_2017} are left for future work. In the bottom panel of Fig.~3b, we plot the Noise Visibility Ratio, a proxy for noise temperature, which is the ratio between the noise power spectral density with the pump on and the pump off ($\rm{NVR} = P_{\rm{on}}/P_{\rm{off}}$). The NVR of our device was comparable to that of JRM-based quantum-limited amplifiers measured in the same system~\cite{bergeal_phase-preserving_2010}, showing that the SNAIL can potentially be used as a quantum-limited amplifier as well.

In future work, the dipole nature of the SNAIL enables the replacement of a single SNAIL with an array of SNAILs (Fig.~4a) for increased dynamic range or bandwidth. This strategy is directly analogous with the tessellation of DC-SQUIDs in the JPA~\cite{castellanos-beltran_amplification_2008} or the tessellation of transmission line unit cells in the Josephson traveling wave parametric amplifier~\cite{macklin_nearquantum-limited_2015}, both four-wave mixing amplifiers. The SNAIL allows this tessellation for three-wave mixing amplifiers~\cite{zorin_josephson_2016}.

Another relevant application of the SNAIL is for Kerr-free tunable coupling between two cavities (Fig.~4b). By pumping at the difference frequency between the red and blue modes, the three-wave mixing capabilities of the SNAIL generate the effective interaction $g_{\rm{eff}}(t) (a_b^\dagger a_r + a_r^\dagger a_b)$. Such tunable couplings have been used as tools for quantum optics in the microwave regime~\cite{flurin_superconducting_2015, sliwa_reconfigurable_2015}, however previous schemes always required that the red and blue modes inherit self-Kerr and cross-Kerr terms from the nonlinearity~\cite{pfaff_schrodingers_2016}. In principle, using a SNAIL with $c_4 = 0$ allows one to generate the same effective interaction without the harmful Kerr terms.

In conclusion, we have introduced a dipole realization of a pure $\bm{\varphi}^3$ nonlinearity, the SNAIL, which performs three-wave mixing in the microwave regime of superconducting quantum circuits. We have also demonstrated the SNAIL's three-wave mixing capabilities by integrating it into a non-degenerate three-wave mixing amplifier, whose performance matched that of similar quantum-limited amplifiers based on the JRM. However, the SNAIL has the advantage of supporting parametric three-wave processes without introducing extra Kerr-induced frequency shifts. Left for future work is the demonstration of this feature. Furthermore, the dipole nature of the SNAIL allows for the integration of cubic nonlinearity wherever the circuit designer deems necessary. Specifically, this enables the construction of three-wave mixing circuits that share the work between many junctions to increase bandwidth and input power handling capabilities, without the addition of prohibitively large DC loops. Thus, the SNAIL will likely be at the heart of the next generation of robust quantum-limited parametric amplifiers and converters.

We acknowledge fruitful discussions with Clarke Smith and Luigi Frunzio. We also acknowledge the Yale Quantum Institute. Facilities use was supported by YINQE, the Yale SEAS cleanroom, and NSF MRSEC DMR 1119826. This research was supported by ARO under Grant No. W911NF-14-1-0011 and W911NF-16-10349, AFOSR under Grant No. FA9550-15-1-0029 and by MURI-ONR Grant No. N00014-16-1-2270.


%

\end{document}